\documentclass[twocolumn,nofootinbib]{revtex4}

\usepackage{blindtext,mathtools,amsmath,amsfonts,amssymb,physics,slashed,esvect,mathrsfs,dsfont,bbold}
\usepackage[utf8]{inputenc}
\usepackage{graphicx}
\usepackage{float}
\usepackage[footnotesize]{caption}

\usepackage{subfig}

\usepackage{multirow}

\usepackage{empheq}

\usepackage{pstricks,pst-node}

\DeclareCaptionJustification{justified}{\leftskip=0pt \rightskip=0pt \parfillskip=0pt plus 1fil}

\usepackage{xcolor}

\mathchardef\Re="023C 
\mathchardef\Im="023D

\pdfoutput=1

\begin{document}
\title{Dark energy explained by a bias in the measurements}

\author{Vincent Deledicque}
\email[]{vincent.deledicq@gmail.com}
\affiliation{No affiliation}

\date{\today}

\begin{abstract}
Typical cosmological models are based on the postulate that space is homogeneous. Space however contains overdense regions in which matter is concentrating, leaving underdense regions of almost void. The evolution of the scale factor of the universe has been established from measurements on SNIa. Since such events occur in regions were matter is present, we may expect that most of the SNIa are located in overdense regions. This means that the evolution of the scale factor has been established in a biased manner, by considering only information coming from overdense regions, excluding the one from the underdense regions. We develop a simple model to analyze the effect of this bias, and show that it leads to the appearance of a new tensor in the Einstein equation of general relativity, which can account for the apparent acceleration of the expansion of the universe. We further show that this tensor tends to be proportional to the FLRW metric tensor, and that the constant of proportionality quantitatively corresponds to the measured cosmological constant with a remarkable accuracy. We finally explain why these properties remain valid for other techniques used in determining the dynamics of the universe, such as the baryon acoustic oscillations.
\end{abstract}

\maketitle


\section{Introduction}

Typical cosmological models are based on the postulate that the universe is homogeneous and isotropic in its spatial dimensions. This postulate is generally known as the Cosmological principle. Obviously, at small scales, space presents heterogeneities and anisotropies, but we take the Cosmological principle to apply only on the largest scales, where local variations are averaged over. The homogeneity and isotropy of space at such scales imply that it would be maximally symmetric, leading to the well-known Friedmann-Lema\^\i tre-Robertson-Walker (FLRW) metric. In typical cosmological models, this metric is then used to determine the left part of the Einstein equation of general relativity:
\begin{eqnarray} \label{Einsteinequation}
	G_{\mu\nu} + \Lambda g_{\mu\nu} = 8\pi G T_{\mu\nu}\,,
\end{eqnarray}
where $\Lambda$ is the cosmological constant. The right part is determined by estimating the average stress-energy tensor of all identified sources. Solving the Einstein equation leads finally to the Friedmann equations, allowing to predict the evolution of the scale factor $a(t)$ of the universe in function of its density. In simple words, this approach allows to predict the global evolution of the universe in function of its content.

It is known since the beginning of the study of cosmology that space is not perfectly homogeneous and isotropic, but the effects of this characteristic on the evolution of the universe have been investigated seriously since some decades only. In particular, several authors have postulated that the consideration of the inhomogeneity of space could account for the accelerating expansion of the universe, as evidenced initially by \cite{Riess} and \cite{Perlmutter} from the observations of distant Type Ia supernovae (SNIa) and further confirmed over the years, when more and more results were obtained, see \cite{Scolnic}.

First of all, important efforts have been put by several authors on the investigation of back-reaction effects, see \cite{Buchert}, \cite{Kolb}, \cite{Clifton}, and many others. In such studies, the accelerating expansion of the universe is explained by the fact that the Einstein equation of general relativity should not be applied as such for the FLRW metric, but should be averaged in some way to take into account the inhomogeneous reality. The non-commutativity of the averaging procedure would lead to new terms in the averaged Einstein equation that would account for the observed acceleration.

In a different approach, other authors suggested that our galaxy is located in a large underdense region, and that this would alter the observations of SNIa in such a way that it would lead to an apparent accelerating expansion of the universe, see for example \cite{Iguchi}, \cite{Ishak} and \cite{Alexander}. While in the first explanation, this accelerating expansion is considered as being a real behavior, in this second explanation, it would just be an illusion.

We investigate here a different effect related to the inhomogeneity of space. Reminding that this inhomogeneity is due to a non-homogeneous distribution of matter (matter is mainly concentrating in overdense regions, leaving other underdense regions of almost void), we would expect that SNIa do not occur randomly over space. In fact, the more matter in a region, the more stars can form there, and the more chances we have to observe a SNIa. We could thus fear that most of the SNIa that have been observed to establish the evolution of the scale factor are located in overdense regions. If this was the case, that would mean that we are establishing the evolution of the scale factor in a biased manner, by considering only information coming from these overdense regions, and excluding the one from the underdense regions. At a first sight, we could believe that such a bias would only have a slight effect and could in particular not explain the observed accelerated expansion of the universe. We will however show the contrary. 

To further illustrate the interest to investigate the effect of this bias, let us note an intriguing coincidence. Observations performed during the two last decades to determine the dynamics of the universe lead to conclude that a non-zero cosmological constant exists in Eq.\ $(\ref{Einsteinequation})$. Applying this equation to the universe by assuming that it is characterized by the FLRW metric tensor and that space is flat, we typically write its first diagonal component as
\begin{equation}\label{qh2}
	3\frac{\dot{a}^2}{a^2} = 8\pi G \left(\overline{\rho} + \rho_\Lambda\right)\,,
\end{equation}
where $\overline{\rho}$ is the average density of the matter content, and where the cosmological constant contribution has been converted in terms of a density $\rho_\Lambda$. Written as such, the right hand side of Eq.\ $(\ref{qh2})$ hence implies the existence of two distinct driving forces for the dynamics, a first one related to matter, and a second one related to a distinct and still unknown concept, called dark energy, for want of anything better.

Current values estimated from the measurements for the density parameters are $\Omega_M \approx 0.27$ for matter and $\Omega_\Lambda \approx 0.73$ for dark energy (see for example \cite{Farooq}), meaning thus that $\rho_\Lambda \approx 0.73/0.27\, \overline{\rho} \approx 2.7\, \overline{\rho}$.
Quantitatively, this means that for the current time, Eq.\ $(\ref{qh})$ can be written as
\begin{equation}\label{qh}
	3\frac{\dot{a}^2}{a^2} = 8\pi G \left(3.7\, \overline{\rho}\right)\,,
\end{equation}
in other words, that globally, the dynamics is governed by an overall density corresponding to 3.7 times the current average density of matter.

It happens that quantitatively this overall density corresponds to the average density of matter in overdense regions. Indeed, from the detailed estimations made by \cite {Cautun}, we know that overdense regions occupy 23\% of the universe's volume, and contain 85\% of the universe's total mass. Since density is directly proportional to the mass, and inversely proportional to the volume, we deduce that the density in the overdense regions $\rho_{(o)}$ is
\begin{equation}
	\rho_{(o)} \approx \frac{0.85}{0.23}\, \overline{\rho} \approx 3.7\, \overline{\rho}\,.
\end{equation}
Of course, this could be a simple coincidence, but the aim of this article is to demonstrate that this perfect match has a real physical origin, related to the bias in the measurements.

In section $\ref{S1}$ we develop a simplified model to represent the inhomogeneous reality of space, by considering only two kinds of regions, namely overdense and underdense regions. On the basis of this model, we then deduce how the bias influences the observed dynamics. In particular, we show that it leads to a supplementary term in the Einstein equation of general relativity. In section $\ref{S2}$ we investigate this new term, and demonstrate that it tends to be proportional to the FLRW metric tensor. In section $\ref{S3}$, we then briefly discuss the theory, in particular its application to other approaches that have been used to determine the dynamics of the universe.


\section{A simple model to predict the effect of the bias}\label{S1}

From now on we assume that the cosmological constant $\Lambda$ is zero. The Einstein equation of general relativity hence reads
\begin{equation} \label{Einsteinequation2}
	G_{\mu\nu} = 8\pi G T_{\mu\nu}\,.
\end{equation}

The simplest approach that can be considered to model the universe is the one in which matter is distributed in a perfectly homogeneous way. Such an assumption is the one that has been used to derive the Friedmann equations, but it is unable to distinguish overdense and underdense regions, and it is thus inappropriate for an investigation of the effects of the bias identified above. A slightly more elaborated representation of the universe would consider that space is made only of two kinds of regions, namely homogeneous overdense regions and homogeneous underdense regions, and this is the one we propose to use. Similarly as for the FLRW approach, both sub-regions are characterized by uniform parameters coming from some averaging process over the corresponding sub-region. Obviously, this still does not correspond to reality, but it is yet slightly closer to it with respect to the largely used perfectly homogeneous and isotropic FLRW approach. More advanced theories on inhomogeneities could be used, but as will be shown, the simplified approach we propose here is sufficient to have a conceptual understanding of the issue.

According to Einstein's law of general relativity, overdense regions present a dynamics verifying
\begin{equation}\label{hhq1}
	G_{\mu\nu(o)} = 8\pi G T_{\mu\nu(o)}\,,
\end{equation}
whereas underdense regions present a dynamics verifying
\begin{equation}\label{hhq2}
	G_{\mu\nu(u)} = 8\pi G T_{\mu\nu(u)}\,.
\end{equation}
In these two latter relations we used the subscripts $(o)$ and $(u)$ to specify in which region the tensors are evaluated (overdense and underdense, respectively). On the other hand, it is expected that on the global scale, the universe verifies the following dynamics:
\begin{equation}\label{hhq}
	\overline{G}_{\mu\nu} = 8\pi G \overline{T}_{\mu\nu}\,,
\end{equation}
where $\overline{G}$ and $\overline{T}$ represent, respectively, the Einstein tensor related to the FLRW metric and the average stress-energy tensor.

It should be emphasized that different averaging approaches have been proposed to deduce Eq.\ $(\ref{hhq})$ from Eq.\ $(\ref{hhq1})$ and $(\ref{hhq2})$, some of them leading to an additional term in Eq.\ $(\ref{hhq})$, accounting for the back-reaction effects mentioned in the introduction. The theoretical existence of such a term seems to be largely accepted, however, its significance with respect to the other terms is still under debate, see for example \cite{Buchert2}. Since the aim of this article is to investigate if the bias of the measurements could account for the apparent accelerating expansion of the universe, we will therefore admit that back-reaction effects, if any, are negligible.

Now, Eq.\ $(\ref{hhq})$ describes the dynamics of the universe as a whole, and the Friedman equation that is derived from it is the one we expect verifying when performing measurements on SNIa. The left hand side of Eq.\ $(\ref{hhq})$ can be expressed in terms of the scale factor, which is the parameter that is determined from the SNIa measurements. The right hand side is determined by estimating the average stress-energy tensor of all identified sources. Conceptually, we may consider that the verification then consists in checking that the scale factor $a$ and is temporal evolution indeed leads to an average Einstein tensor that can be related to the average stress-energy tensor by Eq.\ $(\ref{hhq})$, even if in practice the verification is performed differently.

However, since SNIa occur only in the overdense regions (neglecting the contribution of the few ones that could occur in the underdense regions), we determine the dynamics in a biased way, and this has significant consequences. To see what this means in practice, let us define $\Delta G_{\mu\nu(o)}$ and $\Delta T_{\mu\nu(o)}$ such that
\begin{equation}\label{g1}
	G_{\mu\nu(o)} = \overline{G}_{\mu\nu} + \Delta G_{\mu\nu(o)}
\end{equation}
and
\begin{equation}\label{g2}
	T_{\mu\nu(o)} = \overline{T}_{\mu\nu} + \Delta T_{\mu\nu(o)}\,.
\end{equation}
Hence, $\Delta G_{\mu\nu(o)}$ (resp. $\Delta T_{\mu\nu(o)}$) represents the difference between the local Einstein tensor $G_{\mu\nu(o)}$ (resp. the local stress-energy tensor $T_{\mu\nu(o)}$) and the average Einstein tensor $\overline{G}_{\mu\nu}$ (resp. the average stress-energy tensor $\overline{T}_{\mu\nu}$) in overdense regions. Using Eq.\ $(\ref{g1})$ and $(\ref{g2})$, we may write Eq.\ $(\ref{hhq1})$ as
\begin{equation}\label{gq}
	\overline{G}_{\mu\nu} + \Delta G_{\mu\nu(o)} = 8 \pi G\left(\overline{T}_{\mu\nu} + \Delta T_{\mu\nu(o)}\right)\,.
\end{equation}
Also, combining Eq.\ $(\ref{hhq1})$, $(\ref{hhq})$, $(\ref{g1})$ and $(\ref{g2})$, we deduce that
\begin{equation}\label{io}
	\Delta G_{\mu\nu(o)} = 8\pi G \Delta T_{\mu\nu(o)}\,.
\end{equation}
This means that the second terms of the left and right hand sides of Eq.\ $(\ref{gq})$ cancel, and hence that we may use indifferently Eq.\ $(\ref{gq})$ in the overdense region or Eq.\ $(\ref{hhq})$ on the global scale to establish the evolution of the scale factor.

Now, here comes the subtlety. In theory, we expect and hence assume that space is globally perfectly homogeneous and isotropic, presenting everywhere the same average scale factor and the same related average Einstein tensor $\overline{G}_{\mu\nu}$. Importantly is that this assumption is also used when performing measurements on SNIa, meaning in particular that the way we interpret such measurements implicitly implies that we assume that on average, the Einstein tensor corresponds to $\overline{G}_{\mu\nu}$ at the points where SNIa occur (this will be shown more formally thereafter). But since SNIa only occur in overdense regions, it implies in practice that it is not the overall universe that is assumed to be characterized by the average Einstein tensor $\overline{G}_{\mu\nu}$, but instead the overdense regions. In other words, in practice, due to the bias, we implicitly assume that $\Delta G_{\mu\nu(o)}$ vanishes. This means that the dynamics in overdense regions does not verify Eq.\ $(\ref{gq})$ but instead
\begin{equation}\label{fff}
	\overline{G}_{\mu\nu} = 8 \pi G\left(\overline{T}_{\mu\nu} + \Delta T_{\mu\nu(o)}\right)\,.
\end{equation}
This latter equation provides a relation to determine the evolution of the scale factor (on which $\overline{G}_{\mu\nu}$ depends), and even if this equation has been derived from measurements performed in overdense regions only, it is important to remind that the scale factor is not a local property but a global property, representative of the overall space. Now, in Eq.\ $(\ref{fff})$, the second term of the right hand side does not cancel anymore with the second one of the left hand side which has been ignored. It is this term that will account for an apparent dark energy effect, as we will see.

Let us now demonstrate all the claims made above. Globally, the investigation in this article will be performed by using a covariant approach. Though, for practical reasons, at some points we will make use of a specific frame of reference. When doing so, we will however show that the results are frame invariant, and can be put back in a covariant way. The specific frame of reference that we will use is based on the co-moving coordinates, for which $t$ is the cosmological time coordinate, and where $(x,y,z)$ are the spatial Cartesian coordinates.

It will also be useful to write the interval as
\begin{equation}\label{79}
	ds^2 = g_{\mu\nu}dx^\mu dx^\nu = \left(\overline{g}_{\mu\nu} + \Delta g_{\mu\nu}\right)dx^\mu dx^\nu\,,
\end{equation}
where $g_{\mu\nu}$ is the local metric tensor, $\overline{g}_{\mu\nu}$ is the FLRW metric tensor, and $\Delta g_{\mu\nu}$ is the difference between the real local and the FLRW metric tensors. In the specific coordinates considered here, $\overline{g}_{\mu\nu}$ is diagonal: $\overline{g}_{tt} = -1$, $\overline{g}_{ii} = a^2$ and all other components are zero.

We have claimed that in practice we implicitly assume that $\Delta G_{\mu\nu(o)}$ is completely negligible. Let us formally demonstrate that the way SNIa measurements are performed indeed lead to such an assumption. Such measurements consist in redshift and luminosity distance measurements. From redshift measurements, we deduce the scale factor when the SNIa occurred, while from luminosity distance measurements, we deduce the time coordinate at which this SNIa occurred. Performing such measurements on SNIa occurring at different times hence allows us determining the evolution of the scale factor over time. To understand the effect of the bias on these measurements, it is important to remind how we proceed in practice. 

We start with the luminosity distance $d_L$, defined as
\begin{equation}\label{qk}
	d_L^2 = \frac{L}{4\pi F}\,,
\end{equation}
where $L$ is the absolute luminosity emitted by the source (supposed to be known) and $F$ is the flux measured by the observer. The absolute luminosity represents an amount of energy per unit time. This parameter is expressed in function of the proper time of the source. In practice, due to the Cosmological Principle, this proper time is supposed to be identical everywhere and corresponding to the Cosmological time. However, a local perturbation $\Delta g_{tt}$ could exist at the SNIa, and could modify the local proper time. This perturbation should hence theoretically be taken into account to correctly express $L$ in Eq.\ $(\ref{qk})$ and deduce $d_L$ for every specific SNIa. But when performing luminosity distance measurements on a large sample, statistics apply, and if measurements were performed in an unbiased way over whole space, the correction that has to be applied on $L$ would globally cancel, positive perturbations being equivalently compensated by negative perturbations ($\Delta g_{tt}$ cancels on average over space). Due to the bias, however, measurements are performed in the overdense region only, where $\Delta g_{tt}$ does not cancel on average. 

It should be stressed that Eq.\ $(\ref{qk})$ is only valid for perfect standard candles. Small but significant variations of the peak luminosities of SNIa have been observed, meaning hence that SNIa cannot be considered as such. Therefore, in practice, correction terms have to be taken into account to transform them into genuine standard candles. These correction terms represent different effects (such as the influence of the amount of nickel-56 for example), but none of them is supposed to compensate a non-zero average $\Delta g_{tt}$. So when using Eq.\ $(\ref{qk})$, even together with correction terms, by expressing the absolute luminosity in the average FLRW time frame, we implicitly assume that $\Delta g_{tt}$ cancels in the region where SNIa measurements are performed, i.e., in overdense regions. Let us highlight here that it is completely different to assume that $\Delta g_{tt}$ vanishes on average over whole space or that $\Delta g_{tt}$ vanishes on average in the overdense regions only.

We now consider the effect of the bias on the redshift measurements. To fix ideas, let us consider a source (typically a SNIa) emitting light with a known temporal characteristic. A first signal is emitted at time $t_1$ by such a source located at $x = x_1$ and reaches at time $t_2$ an observer located along the $x$ direction at $x = x_2$. A second signal is emitted from the same source at time $t_1 + \Delta t_1$ and reaches the observer at time $t_2+\Delta t_2$. 

Light follows a null geodesic, so
\begin{equation}
	\left(\overline{g}_{tt} + \Delta g_{tt}\right)dt^2 + \left(\overline{g}_{xx}+\Delta g_{xx}\right)dx^2 + 2\Delta g_{xt}dxdt = 0\,.
\end{equation}
At each point, $\Delta g_{xx}$ and $\Delta g_{xt}$ could be everything, but we will show than in practice we assume that they cancel in overdense regions. Solving for $dx/dt$, we find
\begin{equation}
	\frac{dx}{dt} = \frac{-2\Delta g_{xt} \pm \sqrt{\delta}}{2(\overline{g}_{xx}+\Delta g_{xx})}\,,
\end{equation}
where
\begin{equation}
	\delta = 4(\Delta g_{xt})^2 - 4(\overline{g}_{xx} + \Delta g_{xx})\left(\overline{g}_{tt} + \Delta g_{tt}\right)\,.
\end{equation}
We have one solution for a signal traveling in the positive direction, and one solution for a signal traveling in the negative direction. Since space is expected to be isotropic, however, this equation should on average provide a similar result in magnitude for both signals, but with only a change of sign. This is only possible if on average $\Delta g_{xt}$ vanishes. Then, replacing $\overline{g}_{tt} = -1$ and $\overline{g}_{xx} = a^2$, we get
\begin{equation}\label{pp1}
	\sqrt{\frac{1-\Delta g_{tt}}{a^2+\Delta g_{xx}}}dt = \pm dx\,.
\end{equation}
We now integrate Eq.\ $(\ref{pp1})$ along the $x$ direction. For the signal emitted at $t_1$, we get
\begin{equation}
	\int_{t_1}^{t_2} \sqrt{\frac{1-\Delta g_{tt}}{a^2+\Delta g_{xx}}}dt = \pm\int_{x_1}^{x_2}dx\,.
\end{equation}
Considering the equivalent relation for the second signal we show that
\begin{equation}\label{ggqq}
	\int_{t_1}^{t_1+\Delta t_1}\sqrt{\frac{1-\Delta g_{tt}}{a^2+\Delta g_{xx}}}dt = \int_{t_2}^{t_2+\Delta t_2}\sqrt{\frac{1-\Delta g_{tt}}{a^2+\Delta g_{xx}}}dt\,.
\end{equation}
For a small variation in time, the components of the metric may be considered as constant, and we deduce that
\begin{eqnarray}\label{ivc}
	\frac{\sqrt{a^2(t_1)+\Delta g_{xx}(x_1,t_1)}}{\Delta t_1\sqrt{1-\Delta g_{tt}(x_1,t_1)}} = \frac{\sqrt{a^2(t_2)+\Delta g_{xx}(x_2,t_2)}}{\Delta t_2\sqrt{1-\Delta g_{tt}(x_2,t_2)}}\,.
\end{eqnarray}
So, Eq.\ $(\ref{ivc})$ is the one that should be used to determine the scale factor from measurements. However, in practice, we use the following equation:
\begin{equation}\label{pep}
	\frac{a(t_1)}{\Delta t_1} = \frac{a(t_2)}{\Delta t_2}\,.
\end{equation}
By using Eq.\ $(\ref{pep})$ instead of Eq.\ $(\ref{ivc})$, we make an implicit assumption. Let us in particular consider the left hand side of Eq.\ $(\ref{ivc})$. At the source of the signals, $\Delta g_{xx}$ and $\Delta g_{tt}$ could be everything, meaning that we do not know their value for a specific source. But if redshift measurements are carried out for several sources occurring at the same temporal variable $t$, statistics apply. 

If measurements were performed over whole space in an unbiased way, the deviations $\Delta g_{xx}$ and $\Delta g_{tt}$ in overdense regions would globally cancel with the ones of the underdense regions, and, on average, the left hand side of Eq.\ $(\ref{ivc})$ would indeed be equivalent to the left hand side of Eq.\ $(\ref{pep})$. Due to the bias, however, measurements are performed in the overdense region only, where $\Delta g_{xx}$ and $\Delta g_{tt}$ do not cancel on average. We have already shown that we erroneously assume that on average $\Delta g_{tt}$ vanishes in the overdense regions. So by using Eq.\ $(\ref{pep})$ instead of Eq.\ $(\ref{ivc})$, we also erroneously assume that $\Delta g_{xx}$ cancels in overdense regions. In fact, by considering light emitted in different directions, we show in a similar way that all other components $\Delta g_{ij}$ and $\Delta g_{it}$ are assumed to cancel as well.

In summary, in the luminosity and redshift measurements performed in practice, we assume that all components $\Delta g_{\mu\nu}$ vanish in overdense regions, meaning that this tensor is null. This has been demonstrated in a specific frame of reference, but since the tensor is null, this result is frame invariant. The fact that $\Delta g_{\mu\nu}$ is neglected in overdense regions means that the local metric is indeed assumed to be the FLRW metric in those regions and as a consequence that all related tensors (in particular the Einstein tensor $G_{\mu\nu}$) are the ones calculated from the FLRW metric. This demonstrates that in practice we assume that $\Delta G_{\mu\nu(o)}$ vanishes. Also, a consequence thereof is that on the global scale, the universe's average metric tensor cannot be considered as being the FLRW one. This will be explained in more detail in section \ref{S3}.

Let us finally stress the following. The implicit assumption we make in practice on $\Delta G_{\mu\nu(o)}$ has also consequences on $\Delta T_{\mu\nu(o)}$. Indeed, in Eq.\ $(\ref{fff})$, $G_{\mu\nu(o)}$ and $T_{\mu\nu(o)}$ are diagonal tensors in our specific frame of reference, and as a consequence $\Delta T_{\mu\nu(o)}$ has also to be diagonal. If this was not the case, this would formally mean that the assumption made on $\Delta G_{\mu\nu(o)}$ also implies that we implicitly assume that the off-diagonal components of $\Delta T_{\mu\nu(o)}$ vanish. In practice, these off-diagonal are however not considered, since measurements performed only involve the diagonal components of Eq.\ $(\ref{fff})$, leading to the two Friedmann equations.


\section{Link between $\Delta T_{\mu\nu(o)}$ and the FLRW metric}\label{S2}

In the previous section we have shown that the dynamics that we measure in practice is the one described by Eq.\ $(\ref{fff})$. In this section we will further examine the tensor $\Delta T_{\mu\nu(o)}$ and show that it tends to be proportional to the FLRW metric. We will also show that this tensor can be significant. This study will be performed in the specific frame of reference $(t,x,y,z)$, but once again, a covariant conclusion will be deduced from it.

We start by examining the first diagonal component, i.e., $\Delta T_{tt(o)}$. Let us consider a volume $V$ of space sufficiently large so that it can be considered as representative of the universe. In this volume, overdense regions occupy a volume $V_{(o)}$ and have an average density $\rho_{(o)}$, while underdense regions occupy a volume $V_{(u)}$ and have an average density $\rho_{(u)}$. For simplicity, in a first step, we will even consider the limiting case for which $\rho_{(u)} = 0$, meaning thus that all matter is concentrated in the overdense regions, and that underdense regions are empty.

In overdense regions, according to the first diagonal component of the Einstein equation of general relativity, we have
\begin{equation}\label{p}
	G_{tt(o)} = 8\pi G \rho_{(o)}\,,
\end{equation}
whereas in underdense regions we have
\begin{equation}
	G_{tt(u)} = 8\pi G \rho_{(u)}\,.
\end{equation}
The global average scale factor $a$ presents a behavior lying between the ones of those both regions:
\begin{equation}\label{qd}
	\overline{G}_{tt} =  3\frac{\dot{a}^2}{a^2} = 8\pi G \overline{\rho}\,,
\end{equation}
where we assumed a globally flat universe and where $\overline{\rho}$ is the overall average density:
\begin{equation}\label{w1}
	\overline{\rho} = \frac{\rho_{(o)}V_{(o)} + \rho_{(u)}V_{(u)}}{V_{(o)}+V_{(u)}} = \rho_{(o)}\frac{V_{(o)}}{V_{(o)}+V_{(u)}}\,,
\end{equation}
since we assumed that $\rho_{(u)} = 0$. We just rewrote the Eq.\ $(\ref{hhq1})$ to $(\ref{hhq})$ for the first diagonal component in our specific frame of reference.

Using Eq.\ $(\ref{w1})$, we may write Eq.\ $(\ref{p})$ as
\begin{equation}\label{qkz}
	G_{tt(o)} = 8\pi G\overline{\rho}\frac{V_{(o)}+V_{(u)}}{V_{(o)}} = 8\pi G\left(\overline{\rho} + \rho_\Lambda\right)\,,
\end{equation}
where we defined
\begin{equation}\label{w2}
	\rho_\Lambda = \overline{\rho}\frac{V_{(u)}}{V_{(o)}}\,.
\end{equation}
Moreover, we demonstrated that in practice we neglect $\Delta G_{\mu\nu(0)}$. This means that we pretend that $G_{tt(o)} = \overline{G}_{tt}$, and due to this assumption, Eq.\ $(\ref{qkz})$ reads
\begin{equation}\label{yu}
	\overline{G}_{tt} = 8\pi G\left(\overline{\rho} + \rho_\Lambda\right)\,.
\end{equation}
This equation determines the apparent dynamics that we will observe in practice. Comparing Eq.\ $(\ref{yu})$ with Eq.\ $(\ref{fff})$, we deduce that in our specific frame of reference, the first diagonal component of $\Delta T_{\mu\nu(o)}$ corresponds to
\begin{equation}
	\Delta T_{tt(o)} = \rho_\Lambda\,.
\end{equation}

Next, we consider the three other diagonal terms of $\Delta T_{\mu\nu(o)}$, which are identical for isotropic reasons. Since the first diagonal component has been written as a density, we will write the other ones in terms of a pressure $p_\Lambda$, such that
\begin{equation}\label{aq}
	\Delta T_{\mu\nu(o)} = \left(
	\begin{array}{c c c c}
		\rho_\Lambda & 0 & 0 & 0\\
		0 & a^2 p_\Lambda & 0 & 0\\
		0 & 0 & a^2 p_\Lambda & 0\\
		0 & 0 & 0 & a^2 p_\Lambda
	\end{array} \right)\,.
\end{equation}
Now, as $\nabla^\mu T_{\mu\nu} = \nabla^\mu \overline{T}_{\mu\nu} = 0$, we deduce that we also have $\nabla^\mu \left(\Delta T_{\mu\nu(o)}\right) = 0$, meaning that $T_{\mu\nu(o)}$ is related to a continuity equation. If we can establish such a relation, we will be able to deduce the pressure $p_\Lambda$. We therefore differentiate Eq.\ $(\ref{w2})$ with respect to the cosmological time coordinate:
\begin{equation}\label{r}
	\dot{\rho}_\Lambda = \dot{\overline{\rho}}\frac{V_{(u)}}{V_{(o)}} + \overline{\rho}\frac{\dot{V}_u}{V_{(o)}} - \overline{\rho}\frac{V_{(u)}\dot{V}_o}{V_{(o)}^2}\,.
\end{equation}
We may then replace $\dot{\overline{\rho}}$ by using the continuity equation for $\overline{\rho}$. For a matter dominated universe, this one reads
\begin{equation}
	\dot{\overline{\rho}} = -3\frac{\dot{a}}{a}\overline{\rho}\,,
\end{equation}
so we can write Eq.\ $(\ref{r})$ as
\begin{equation}
	\dot{\rho}_\Lambda = -3\frac{\dot{a}}{a}\overline{\rho}\frac{V_{(u)}}{V_{(o)}} + \overline{\rho}\frac{V_{(u)}}{V_{(o)}}\left(\frac{\dot{V}_u}{V_{(u)}} - \frac{\dot{V}_o}{V_{(o)}}\right)\,.
\end{equation}
This latter equation can then be written in the form of a continuity equation
\begin{equation}\label{f}
	\dot{\rho}_\Lambda = -3\frac{\dot{a}}{a}\left(\rho_\Lambda + p_\Lambda\right)
\end{equation}
if we define
\begin{equation}\label{uu}
	p_\Lambda = -\frac{a}{3\dot{a}}\rho_\Lambda\left(\frac{\dot{V}_u}{V_{(u)}} - \frac{\dot{V}_o}{V_{(o)}}\right)\,.
\end{equation}

The expression in parentheses is a priori time dependent, and cannot be simplified in a general way: it depends on how fast inhomogeneities are developing. However, a limiting value to this expression can be deduced. If initially matter was distributed in an almost homogeneous way, small disturbances appeared and led to the development of inhomogeneities. The driving force of this phenomenon was the gravitational attraction of regions having a slightly larger density than other regions. Matter from those latter regions were then progressively attracted towards the former ones, leaving behind them regions of voids that expanded over time. Matter grouped together and formed gravitationally bound structures, such as galaxies and even larger structures. It is a fact however that when matter is gravitationally bound, its volume does not change anymore over time: on the one hand, a stable structure has been reached and inhomogeneities do not further develop to shrink this one. On the other hand, the expansion of the universe does not imply an expansion of such structures, because such effect is compensated by the gravitational attraction. This can be illustrated by galaxies which have stable dimensions over time. Then, when all matter has been grouped together to form gravitationally bound structures, the volume $V_{(o)}$ will not change anymore. This means that at large times we will have $\dot{V}_{(o)} \rightarrow 0$. Also, since at that limit $V_{(o)}$ will not grow anymore, whereas $V_{(u)}$ and $V$ will, $V_{(u)} = V - V_{(o)}$ will progressively tend to $V$, meaning that $\dot{V}_{(u)}/V_{(u)} \rightarrow \dot{V}/V = 3\dot{a}/a$. Using this result in Eq.\ $(\ref{uu})$ we get
\begin{equation}
	p_\Lambda \rightarrow -\rho_\Lambda\,.
\end{equation}
So, if generally $p_\Lambda$ differs from $\rho_\Lambda$, over time, it will tend to the same value in magnitude, but with the opposite sign. As a result, at this limit, Eq.\ $(\ref{aq})$ becomes
\begin{equation}
	\Delta T_{\mu\nu(o)} \rightarrow
	-\frac{\Lambda}{8\pi G}
	\left(
	\begin{array}{c c c c}
		-1 & 0 & 0 & 0\\
		0 & a^2 & 0 & 0\\
		0 & 0 & a^2 & 0\\
		0 & 0 & 0 & a^2
	\end{array} \right)
	\,,
\end{equation}
where
\begin{equation}
	\Lambda = 8\pi G \rho_\Lambda.
\end{equation}
As the last tensor corresponds to the FLRW metric tensor written in our specific frame of reference, and since $\Delta T_{\mu\nu(o)}$ is also a tensor, necessarily, $\Lambda$ must be a scalar. Written in a covariant form, we thus have
\begin{equation}
	\Delta T_{\mu\nu(o)} \rightarrow-\frac{\Lambda}{8\pi G} \overline{g}_{\mu\nu}\,.
\end{equation}
It is important to stress that even if the analysis has been performed in a specific frame of reference, since we could write the result in a covariant form, this one is frame invariant. Introducing this result in Eq.\ $(\ref{fff})$, we get
\begin{equation}\label{uuu}
	\overline{G}_{\mu\nu} + \Lambda \overline{g}_{\mu\nu} = 8 \pi G\overline{T}_{\mu\nu}\,.
\end{equation}
So we have been able to write the observed dynamics by making appear an apparent cosmological constant.

Let us now interpret the new term defined by Eq.\ $(\ref{w2})$. At the early ages of the universe, matter was distributed in a more homogeneous way than currently. Underdense regions were much smaller with respect to overdense regions, meaning that $V_{(u)}/V_{(o)} \ll 1$. Hence, according to Eq.\ $(\ref{w2})$ we have $\rho_\Lambda \ll \overline{\rho}$. This implies that in the parentheses of Eq.\ $(\ref{yu})$, the dominating term is the first one, and the evolution law reduces to the expected one, i.e., without an apparent cosmological constant. But over time, large-scale structures develop, and the matter distribution presents more and more inhomogeneities, meaning that $V_{(u)}/V_{(o)}$ increases. In the parentheses of Eq.\ $(\ref{yu})$, the second term increases progressively, and at some time, when $V_{(u)} > V_{(o)}$, becomes the dominating one. If this term reaches a constant value over time, an apparent acceleration in the evolution of the scale factor will be observed. Theoretically, $\rho_\Lambda$ is time dependent, but we can justify that $\rho_\Lambda$ will indeed tend to a constant value. As explained above, at large times, $V_{(o)}$ will tend to a constant volume, whereas $V_{(u)}$ will tend to $V$. Since $V$ is proportional to $a^3$, $V_{(u)}$ will present the same dependency. On the other hand, $\overline{\rho}$ is proportional to $a^{-3}$. This means that in Eq.\ $(\ref{w2})$, the expression in the right hand side indeed tends to a constant value.

We finally also provide a numerical estimation of $\rho_\Lambda$ to convince us that the proposed theory predicts a cosmological constant in agreement with the measured one. From Eq.\ $(\ref{w2})$ we deduce that
\begin{equation}
	\frac{\rho_\Lambda}{\overline{\rho}} = \frac{V_{(u)}}{V_{(o)}}\,.
\end{equation}
It is generally estimated that at the present time, cosmic voids account for around 80\% of space (see for example \cite {Cautun}, \cite{Tavasoli} and \cite{Falck}). We then get a predicted value of $\rho_\Lambda/\overline{\rho} \approx 4$. On the other hand, as explained in the introduction, we have a measured value of $\rho_\Lambda/\overline{\rho} \approx 2.7$. Despite the difference between predicted and measured values, they present the same order of magnitude. To perform a more accurate comparison, we need to slightly improve our model by taking into account the mass in the underdense regions, which had been neglected up to now. By doing so, we show that
\begin{equation}\label{w4}
	\rho_\Lambda = \overline{\rho}\left(1 - \frac{\rho_{(u)}}{\overline{\rho}}\right)\frac{V_{(u)}}{V_{(o)}}\,.
\end{equation}
We will also use more accurate estimations for the different parameters. According to \cite {Cautun}, voids occupy currently 77\% of the total volume, and contain about 15\% of the total mass. We hence deduce that overdense regions occupy 23\% of the total volume, and contain 85\% of the total mass. The fact that voids contain about 15\% of the total mass means that
\begin{equation}
	0.15 = \frac{\rho_{(u)} V_{(u)}}{\rho_{(o)} V_{(o)} + \rho_{(u)} V_{(u)}} = \frac{\rho_{(u)}}{\overline{\rho}}\frac{V_{(u)}}{V_{(o)} + V_{(u)}}\,,
\end{equation}
from which we deduce that
\begin{equation}
	\frac{\rho_{(u)}}{\overline{\rho}} = 0.15\left(1 + \frac{V_{(o)}}{V_{(u)}}\right) = 0.195\,.
\end{equation}
From Eq.\ $(\ref{w4})$ we then get
\begin{equation}
	\frac{\rho_\Lambda}{\overline{\rho}} = 2.695\,.
\end{equation}
The predicted value of 2.695 is very close to the measured one of 2.7. We can now also better understand the perfect match illustrated in the introduction. In fact, when comparing Eq. $(\ref{p})$ and $(\ref{yu})$, we clearly see that $\rho_\Lambda$ simply represents the difference between the density of the overdense regions with the average one of global space. Adding hence $\rho_\Lambda$ and $\overline{\rho}$ leads to the density of the overdense regions.

It is quite remarkable that the rough approximation we considered, which represents the universe with only two distinct uniform regions and uses macroscopic variables only, is able to predict the value of the apparent dark energy density with such an accuracy. Obviously, the values for $\Omega_\Lambda$ as well as for the volumes and densities of the overdense and underdense regions are known with some uncertainties, which could affect somehow the perfect agreement between prediction and measurements. But the important point to notice is that, despite a possible correction coming from those uncertainties, the model was shown to be able to predict a correct order of magnitude for the apparent dark energy density, and this further supports the proposed explanation.


\section{Discussion}\label{S3}

The theory that has been developed above claims that the determination of the evolution of the scale factor $a$ is biased when this parameter is measured from the observation of SNIa, because such events are not randomly distributed over space. Assuming that matter is distributed homogeneously at large scales through space makes sense to derive some general theoretical concepts, but such a modeling is too simplified on a practical point of view when interpreting results from measurements performed to establish the evolution of the scale factor. If the dynamics of the universe as a whole can be considered to obey the Friedmann equation, the inhomogeneity in the distribution of matter implies that subregions of the universe have their own dynamics. Since observations are only performed in some of those subregions, the evolution of the scale factor is determined in a biased way, and if caution is not taken to interpret the measurements, the evolution derived from them cannot be considered as representative of the global dynamics of the universe. In some sense, the used method suffers from an equivalent Malmquist bias: measurements are performed on events that can be observed, and all regions of the universe that are not 'seen' by this method do not contribute to the as measured average behavior of the universe. 

However, it has to be noticed that the evolution of the scale factor has also been determined by other approaches than the one based on SNIa, and apparently all of them provide similar results. For example, another approach is based on the theory of baryon acoustic oscillations (BAO). This consists in measuring the spatial distribution of galaxies (or other objects) from which a standard ruler is deduced: there is a preference for pairs of galaxies to be separated by a typical distance, and the evolution of the scale factor is determined by measuring the evolution of this typical distance. A priori, this approach is completely independent from the one based on the observation of SNIa, and the fact that it has led to similar results has strengthened the confidence we had in the conclusions drawn from this latter approach.

Yet, we can convince us the BAO approach suffers from the same bias. Indeed, in practice, measurements in this approach imply redshift measurements to deduce radial distances, but also measurements of redshift differences, from which the Hubble expansion rate is calculated. And in a similar way as for the measurements on SNIa, the relations used therefore have been obtained by implicitly assuming that perturbation terms with regard to the FLRW metric vanish. This is obvious: since we consider that at large scales the universe presents the topology of the FLRW metric, there was no need to consider such perturbation terms. However, this has important consequences in practice since, here also, measurements are performed on objects which can be observed, and which are not randomly distributed over space but are preferentially located in overdense regions. This thus means that once again, we assume that it is those regions that are characterized by an average FLRW metric, and hence that their dynamics verifies Eq.\ $(\ref{uuu})$. 

Normally, this should be sufficient to convince us that indeed, the bias in the BAO approach has exactly the same effect as in the SNIa approach. But going further into our reasoning, we could object that there is a major difference between the SNIa approach and the BAO approach: in the first one, we observe punctual events, whereas in the second one, we measure a one-dimensional parameter, i.e., the standard ruler. In the former case, since most SNIa are located in the overdense regions, we easily understand that we measure the local dynamics of that region, but in the latter case, we could believe that we are not affected by the bias: we indeed perform measurements on objects located mainly in overdense regions, but those measurements serve to determine a standard ruler that is so large (much larger than the characteristic dimensions of the overdense regions formed by the galaxies) that it should cross overdense as well as underdense regions and should hence be unbiased, or in other words that it should be representative of the global universe's dynamics. Let us show that this objection is incorrect, and in particular that even if the bias is localized in a particular region, it has global consequences.

By implicitly assuming that in overdense regions the metric tensor exactly corresponds to the FLRW metric (and not to a perturbed one), at the same time we also implicitly assume a shift in the average metric of the global universe. Indeed, since this latter metric corresponds to some kind of average between the metrics of the overdense and underdense regions, in the general case it cannot be equal to one of them. So the bias in the BAO measurements also implies that the metric of the global universe does not correspond to the FLRW metric. As a consequence, the Einstein tensor of the global universe is not equal to the one corresponding to the FLRW metric. It has been modified due to the bias. The way it has been modified is obviously related to how the bias has affected the metric in the overdense regions, and this is simple: the dynamics measured in that region is such that the FLRW metric verifies Eq.\ $(\ref{uuu})$. But even if this equation has been deduced specifically in the overdense region only, we should keep in mind that the FLRW metric is not a local property of the overdense region itself, this tensor having no spatial dependence. Contrary to the local metric, the FLRW tensor remains the same through whole space. This means that, as a consequence of the bias, the Einstein tensor of the global universe is modified so that an identical behavior of the average FLRW metric is obtained. Therefore, it becomes
\begin{eqnarray}
	\overline{G}_{\mu\nu} \rightarrow \overline{G}_{\mu\nu} + \Lambda \overline{g}_{\mu\nu}\,,
\end{eqnarray}
and so the dynamics of the global universe as predicted by the BAO approach verifies Eq.\ $(\ref{uuu})$ as well. More generally, all approaches having the same bias will reach an identical conclusion, whatever they are.

Let us finally note that, since several years, a tension has been highlighted between several methods, meaning that the as measured evolution of the scale factor (and the related parameters, in particular the Hubble constant) differ between those methods in a non-negligible way (see \cite{Riess2}). In the spirit of the proposed theory, this tension could possibly be explained by the fact that the bias could slightly differ between those methods. The volumes occupied by the overdense and underdense regions are not well-defined, and the phenomena or the objects that are observed could possibly be related to slightly different overdense regions. Moreover, the BAO approach pays in its application a particular attention to slightly denser regions than the average overdense regions, and this could also alter the effect of the bias. At this stage, such explanations are however purely hypothetical, and it is not the aim of this article to further investigate them.


\section{Conclusion}

We first noticed that measurements performed to establish the evolution of the scale factor of the universe may contain a bias due to the fact that SNIa are not randomly distributed over space but instead occur preferentially in overdense regions. We developed a model to investigate the effects of this bias on the apparent evolution of the scale factor. This is a simplified model that considers only two kinds of regions, namely homogeneous overdense and homogeneous underdense regions. On the basis of this model, we showed that the bias was responsible for the appearance of a new term in the Einstein equation of general relativity, and that this term tend to be proportional to the metric tensor. We showed that the constant of proportionality predicted by the model is in good agreement with the measured cosmological constant. We further explained why this theory was applicable for other approaches that have been used to determine the evolution of the scale factor, such as the baryon acoustic oscillations. We hence conclude that the bias could account for the apparent accelerated expansion of the universe, without needing the dark energy assumption.  



\begin{thebibliography}{99}
	\bibitem{Riess}
	A. G. Riess et al. (Supernova Search Team Collaboration), Astron. J. {\bf 116} (1998) 1009
	
	\bibitem{Perlmutter}
	S. Perlmutter et al. (Supernova Cosmology Project Collaboration), Astrophys. J. {\bf 517} (1999) 565
	
	\bibitem{Scolnic}
	D. M. Scolnic et al., Astrophys. J. {\bf 859} (2018) 101
	
	\bibitem{Buchert}
	T. Buchert, Gen. Relativ. Gravit. {\bf 40} (2008) 467
	
	\bibitem{Kolb}
	E. W. Kolb, Class. Quant. Grav. {\bf 28} (2011) 164009
	
	\bibitem{Clifton}
	T. Clifton, Int. J. Mod. Phys. D {\bf 22} (2013) 1330004
	
	\bibitem{Iguchi}
	H. Iguchi, T. Nakamura and K. Nakao, Prog. Theor. Phys. {\bf 108} (2002) 809
	
	\bibitem{Ishak}
	M. Ishak, J. Richardson, D. Whittington and D. Garred, Phys. Rev. D {\bf 78} (2008) 123531	
	
	\bibitem{Alexander}
	S. Alexander, T. Biswas, A. Notari and D. Vaid, J. Cosmol. Astropart. Phys. {\bf 9} (2009) 025
	
	\bibitem{Farooq}
	O. Farooq, D. Mania and B. Ratra, Astrophys. J. {\bf 764} (2013) 138
	
	\bibitem{Cautun}
	M. Cautun, R. van de Weygaert, B. J. T. Jones and C. S. Frenk, Mon. Not. R. Astron. Soc. {\bf 441} (2014) 2923
	
	\bibitem{Buchert2}
	T. Buchert et al, Class. Quantum Grav. {\bf 32} (2015) 21
	
	\bibitem{Tavasoli}
	S. Tavasoli, K. Vasei and R. Mohayaee,  Astron. Astrophys. {\bf 553} (2013) A15	
	
	\bibitem{Falck}
	B. Falck and M. C. Neyrinck,  Mon. Not. R. Astron. Soc. {\bf 450} (2015) 3239
	
	\bibitem{Riess2}
	A. G. Riess, Nat. Rev. Phys. {\bf 2} (2020) 10
\end{thebibliography}
\bibliographystyle{unsrt}


\end{document}